\documentclass[pra,twocolumn,amsfonts,amssymb,amsmath,showpacs,superscriptaddress,floatfix]{revtex4}
\usepackage{graphicx}
\begin{document}
\title{Cavity QED with Multiple Hyperfine Levels}

\author{K.~M.~Birnbaum}
\altaffiliation[Permanent address: ]{Jet Propulsion Laboratory, California Institute of Technology, Mail Stop 161-135, 4800 Oak Grove Drive, Pasadena, CA 91109, U.S.A.}
\affiliation{Norman Bridge Laboratory of Physics 12-33, California Institute of Technology,
Pasadena, CA 91125, U.S.A.}

\author{A.~S.~Parkins}
\altaffiliation[Permanent address: ]{Department of Physics, University of Auckland, 
Private Bag 92019, Auckland, New Zealand}
\affiliation{Norman Bridge Laboratory of Physics 12-33, California Institute of Technology,
Pasadena, CA 91125, U.S.A.}

\author{H. J. Kimble}
\affiliation{Norman Bridge Laboratory of Physics 12-33, California Institute of Technology,
Pasadena, CA 91125, U.S.A.}

\date{June 8, 2006}

\begin{abstract}
We calculate the weak-driving transmission of a linearly polarized cavity mode strongly coupled to the D2 transition of a single Cesium atom.  Results are relevant to future experiments with microtoroid cavities, where the single-photon Rabi frequency $g$ exceeds the excited-state hyperfine splittings, and photonic bandgap resonators, where $g$ is greater than both the excited- and ground-state splitting.
\end{abstract}


\maketitle

\section{Introduction}

The Jaynes-Cummings model of cavity QED treats an atom as a two-level system.  This is appropriate for a realistic atom when that atom has a cycling transition, typically reached by optical pumping with circularly polarized light \cite{hood98,hood00}.  However, new types of optical resonators such as microtoroids \cite{toroid} and photonic band gap cavities \cite{pbg} do not support circularly polarized modes.  Though these structures with extremely low critical atom and photon numbers show great promise for strong coupling, a more detailed model of the atom must be employed when calculating the properties of these atom-cavity systems \cite{kmb,iqec}.  A linearly polarized mode may couple multiple Zeeman states of the atom.  Additionally, for these very small resonators, the single photon Rabi frequency ($2g$) can be comparable to or larger than the hyperfine splitting of the atom, so that multiple hyperfine levels must be considered when calculating the excitations of the system.  We will consider a linearly polarized single-mode resonator coupled to the D2 ($6S_{1/2} \to 6P_{3/2}$) transition of a single Cesium atom.  However, this may also give some intuition for other multilevel scatterers, such as molecules and excitons \cite{exciton1, exciton2}.

\section{Coupling to Multiple Excited Levels}
\label{sec:toroid}

In order to describe the interaction of the atom with various light fields, it is useful to define the atomic dipole transition operators
\begin{multline}
\label{dipole}
D_{q}(F,F')= \\
\sum_{m_{F}=-F}^{F}|F,m_{F}\rangle \langle F,m_{F}| \mu_{q} |F',m_{F}+q \rangle \langle F',m_{F}+q|
\end{multline}
where $q=\{-1,0,1\}$ and $\mu_{q}$ is the dipole operator for $\{\sigma_{-},\pi, \sigma_{+}\}$-polarization, normalized such that for a cycling transition $\langle \mu \rangle =1$.  We will approximate all atom-field interactions to be dipole interactions.

First, let us consider the case when $g$ is comparable to the hyperfine splitting of the excited states, but still small compared to the ground-state splitting.  This limit is appropriate for the proposed parameters of microtoroid resonators \cite{toroid} and small Fabry-Perot cavities~\cite{hood01}.  If the cavity is tuned near the $F=4 \to F'$ transitions, then the Hamiltonian for the atom cavity system can be written using the rotating wave approximation as
\begin{eqnarray}
\label{H_0}
H_{0} &=& \omega_c a^{\dag} a + \sum_{F'=2}^{5}\omega_{F'}|F'\rangle \langle F'| \nonumber \\
&+&  g\Big(\sum_{F'=3}^{5} a^{\dag}D_{0}(4,F') + D_{0}^{\dag}(4,F')a\Big),
\end{eqnarray}
where $\omega_{F'}$ is the frequency of the $F=4 \to F'$ transition, $\omega_c$ is the frequency of the cavity, and $a$ is the annihilation operator for the cavity mode.  The operator $|F'\rangle \langle F'|$ projects onto the manifold of excited states with hyperfine number $F'$, and may be written more explicitly as $|F'\rangle \langle F'| = \sum_{m_F'} |F',m_F'\rangle \langle F',m_F'|$.  We use units such that $\hbar=1$ and energy has the same dimensions as frequency.  Note that we are treating the cavity as a single-mode resonator with linear polarization.  Fabry-Perot cavities have two modes with orthogonal polarizations, so this model is only appropriate if there is a birefringent splitting which makes one of the modes greatly detuned (compared to $g$) from the atomic resonance.

In the weak-driving limit of an atom-cavity system in the regime of strong coupling, we expect that high transmission will occur when the probe light is resonant with a transition from a ground state of the system to a state in the $N=1$ lowest excitation manifold.  Furthermore, we expect a higher transmission when resonantly exciting an eigenstate which is ``cavity-like,'' i.e., an eigenstate which has larger weight in the field excitation rather than the atomic dipole.

\begin{figure}
	\includegraphics[width=3.3in]{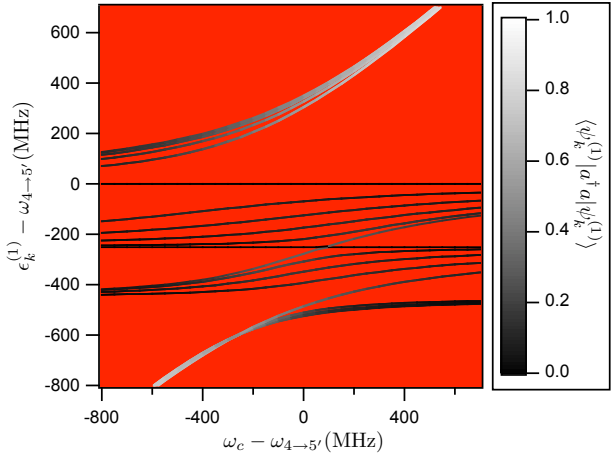}
	\caption[Eigenvalues of a single-mode cavity coupled to Cs $4\to \{3',4',5'\}$]{\label{toroid_eigen} Eigenfrequencies of excitations in the first manifold $\{\epsilon_k^{(1)}\}$ versus cavity detuning.  Color represents $\langle a^{\dag}a\rangle$ for each corresponding eigenvector.  Here $\omega_{4\to 4'}-\omega_{4\to 5'} = -251$~MHz, $\omega_{4\to 3'}-\omega_{4\to 5'} = -452$~MHz \cite{steck}, and $g=450$~MHz \cite{toroid}.  Although there are 36 eigenvectors in the manifold, there are only 20 unique eigenvalues due to symmetry.}
\end{figure}

In Fig.~\ref{toroid_eigen}, we plot the eigenfrequencies $\{\epsilon_k^{(1)}\}$ of $H_0$ determined by the equation $H_0|\psi_k^{(N)}\rangle = \epsilon_k^{(N)} |\psi_k^{(N)}\rangle$.  Here $N$ is the excitation manifold, where $\epsilon_k^{(N+1)}-\epsilon_k^{(N)} \sim \omega_c$ and $\epsilon_k^{(0)}=0$.  Also displayed is $\langle \psi_k^{(1)} | a^{\dag} a |\psi_k^{(1)}\rangle$ for each eigenstate $|\psi_k^{(1)}\rangle$ corresponding to each eigenfrequency $\epsilon_k^{(1)}$, which is a measure of how ``cavity-like'' that state is.  This should give some indication of what cavity and probe detunings yield high transmission.

In order to study the system properties more precisely, we can find the Hamiltonian of the driven system, write the Liouvillian that describes the time-evolution including damping, and calculate the steady state of the system.  We will assume that the cavity resonance is tuned near the $F=4\to F'$ atomic transitions.  We expect that, absent any repumping fields, atomic decays to the $F=3$ ground state will leave the atom uncoupled to the resonator.  To avoid this, we will assume that a classical (coherent-state) driving field tuned near the $F=3\to F'$ transitions is applied to the atom in addition to the probe field which drives the cavity.  In the rotating wave approximation, the Hamiltonian of this driven atom-cavity system in the frame rotating with the cavity probe is
\begin{eqnarray}
\label{H_1}
H_1 &=&\sum_{F'=2}^{5}\Delta_{F'}|F'\rangle \langle F'| + \Delta_r |F=3\rangle\langle F=3| + \Delta_c a^{\dag} a \nonumber \\
&+& g\sum_{F'=2}^{5} \Big(a^{\dag}D_{0}(4,F') + D_{0}^{\dag}(4,F')a\Big) \nonumber \\
&+& \Omega_r \sum_{F'=2}^{5} \Big( D_{0}(3,F') + D_{0}^{\dag}(3,F') \Big)\nonumber \\
&+&\varepsilon a^{\dag} +\varepsilon^{*} a, 
\end{eqnarray}
where $\Delta_{F'} = \omega_{4\to F'} - \omega_p$, $\Delta_r = \omega_r - \omega_{GSS} - \omega_p$, and $\Delta_c = \omega_c-\omega_p$.  Here $\omega_p$ is the probe frequency, $\omega_r$ is the repump frequency, and $\omega_{GSS}\approx 9.2$~GHz is the ground-state splitting of Cs.  The cavity is driven at a rate $\varepsilon$ so that in the absence of an atom the intracavity photon number would be $N_{no~atom} = |\varepsilon|^2 /(\kappa^2 + \Delta_c^2)$, and the atom is driven by the repump field with Rabi frequency $2\Omega_r$.  Here, we have assumed that there is no off-resonant coupling of the cavity mode to the $F=3$ ground states, nor is there off-resonant coupling of the repump light to the $F=4$ states.  We expect that corrections due to those terms will be small when $g,\Omega_r \ll \omega_{GSS}$.

\begin{figure}
	\includegraphics[width=3.3in]{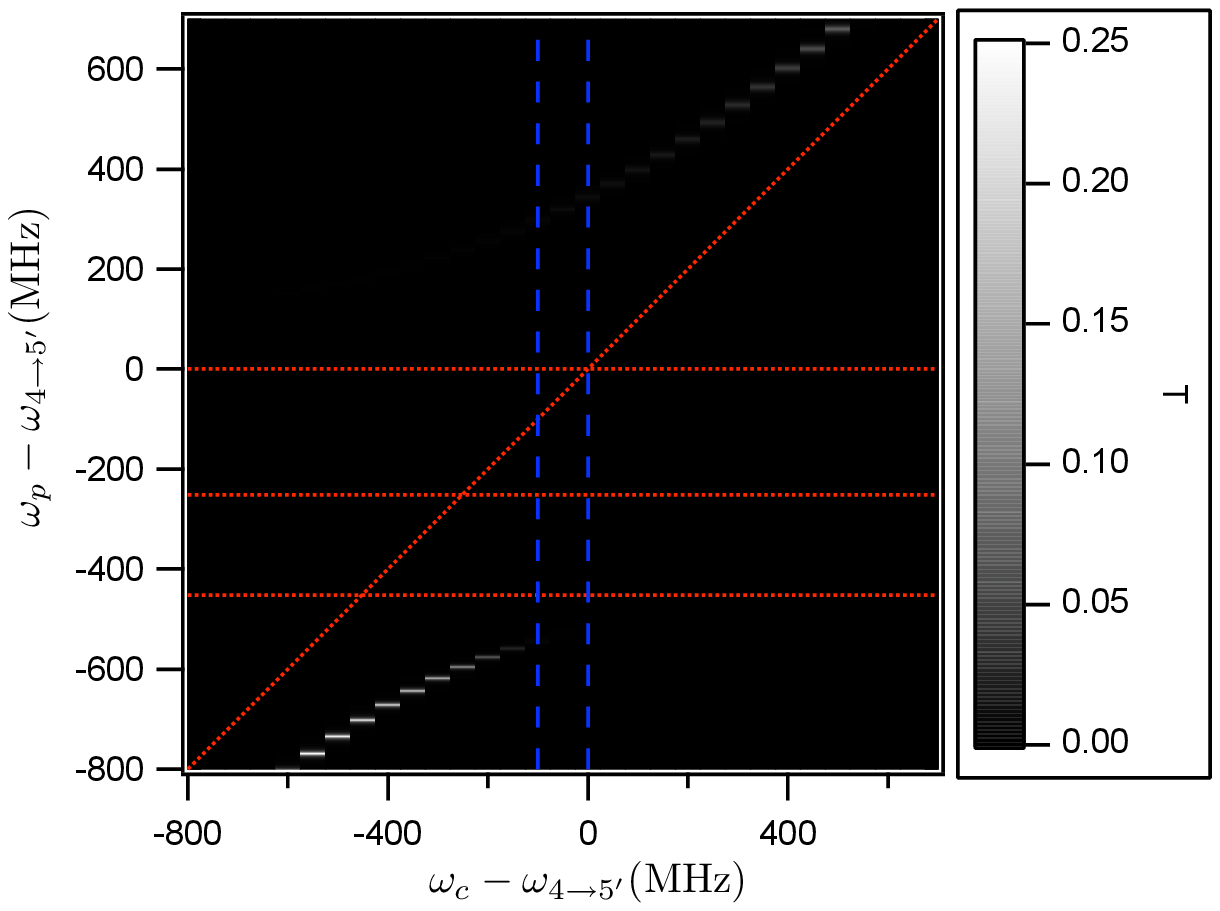}
	\caption[Predicted spectra of a microtoroid coupled to Cs $4\to \{3',4',5'\}$ versus cavity detuning.]{\label{toroidplot} Transmission (intracavity photon number, normalized to the empty cavity on resonance) versus probe and cavity detunings in the weak-driving limit (calculated in a Fock basis of \{0,1\} photons) from Eq.~\ref{master_1}.  Red dotted lines indicate resonances for uncoupled atomic and cavity transitions.  The blue dashed lines indicate the cavity detunings in Figs.~\ref{toroid_slice} and \ref{toroid_4_5}.  Rates are $(g,\kappa,\gamma)=(450,1.75,2.6)$MHz~\cite{toroid},  $\varepsilon = \kappa\gamma/g$, and $\Omega_r = \gamma$.  The repump beam is resonant with the $F=3\to F'=4'$ transition, i.e., $\Delta_r=\Delta_{4'}$.  Values were computed on a grid with a $50$MHz ($0.5$MHz) spacing in $\omega_c$ ($\omega_p$).
	}
\end{figure}

The time evolution of the density matrix $\rho$ of the atom-cavity system is given by the master equation,
\begin{eqnarray}
\label{master_1}
\dot{\rho} = -i[H_1,\rho] + 
\kappa \mathcal{D}[a]\rho +
\gamma \sum_{q,F}\mathcal{D} \Big[\sum_{F'}D_{q}(F,F')\Big]\rho ,
\end{eqnarray}
where $\kappa$ and $\gamma$ are the cavity field and atomic dipole decay rates, respectively,
and the zero-temperature decay superoperator $\mathcal{D}$ acts on the density matrix such that $\mathcal{D}[c]\rho \equiv 2c\rho c^{\dag} - c^{\dag}c\rho -\rho c^{\dag}c$ for any operator $c$.  Note that in the atomic spontaneous emission term (proportional to $\gamma$), we have assumed that all $F'\to F=4$ transitions of the same polarization couple to a common reservoir of vacuum electromagnetic field modes and similarly for all $F'\to F=3$ transitions (but the reservoirs for $F'\to F=4$ and $F'\to F=3$ transitions are independent). This assumption arises from the fact that level shifts due to the atom-cavity coupling will be comparable to the atomic excited state hyperfine splittings (but small compared to the ground state splitting) 
and, therefore, there exists the possibility for coherence, or quantum interference effects between 
transitions of the same polarization from different $F'$ states to a single, common ground-state level
\cite{cardimona82,cardimona83,kmb}. 
Such a possibility is described in the common-reservoir master equation (\ref{master_1}) by
generalized atomic damping terms which couple such transitions.
Note that the choice of independent reservoirs for transitions to the different hyperfine ground states is consistent with our assumption that there is no off-resonant coupling between transitions from different hyperfine ground-state manifolds.  

From the steady-state solution to Eq.~\ref{master_1}, $\dot{\rho}_{ss}=0$, we can compute steady-state expectation values of an operator $c$ by evaluating $\mathrm{Tr}(\rho_{ss}c)$.  We define the normalized cavity transmission $T = \mathrm{Tr}(\rho_{ss}a^{\dag}a) \kappa^2/|\varepsilon|^2$, where $T=1$ for an empty cavity on resonance.  $T$ is plotted in Fig.~\ref{toroidplot} versus cavity and probe detunings.  Notice the similarity to Fig.~\ref{toroid_eigen}, which demonstrates that the qualitative features of the transmission are indeed determined by the eigenvalues and eigenstates of the Hamiltonian.

\begin{figure}
	\includegraphics[width=3.3in]{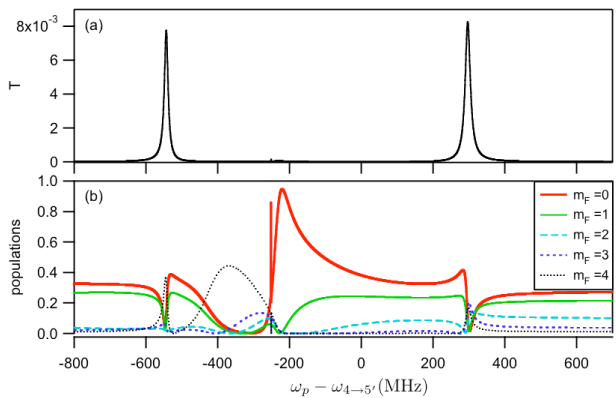}
	\caption[Spectrum and ground-state populations of Cs coupled to a microtoroid.]{\label{toroid_slice} (a) Normalized transmission $T$ and (b) populations in various Zeeman ground states of the $F=4$ manifold as a function of probe detuning, with the cavity frequency fixed at $\omega_c = \omega_{4\to5'}-100$MHz.  Parameters are the same as in Fig.~\ref{toroidplot}.  By symmetry, populations are the same in Zeeman level $m_F$ as in level $-m_F$.  The sharp feature at $\omega_p = \omega_{4\to5'}-251$ is caused by a coherent Raman effect between the probe and the repump light. Note that because of the weak cavity drive and the strong repump light, nearly all of the population is in the $F=4$ manifold.}
\end{figure}

Fig.~\ref{toroid_slice} shows $T$ as a function of probe detuning for fixed cavity frequency along with atomic ground-state populations $\langle F=4,m_F|\rho_{ss}|F=4,m_F\rangle$.  The large swings in the relative populations of various Zeeman states demonstrate the importance of optical pumping in understanding the steady-state behavior of the transmission.  The rapid variation of the populations that occurs near the transmission peaks can be understood by noting in Fig.~\ref{toroid_eigen} that each transmission peak is associated with multiple eigenstates with similar eigenvalues.  These eigenstates have different amplitudes of the Zeeman states and therefore lead to different optical pumping effects.  It should be noted that the width of the transmission peaks are therefore not simply determined by $\kappa$ and $\gamma$ but also by the separation of the various eigenvalues contributing to each peak, making the peaks wider than would be naively expected.

\begin{figure}
	\includegraphics[width=3.3in]{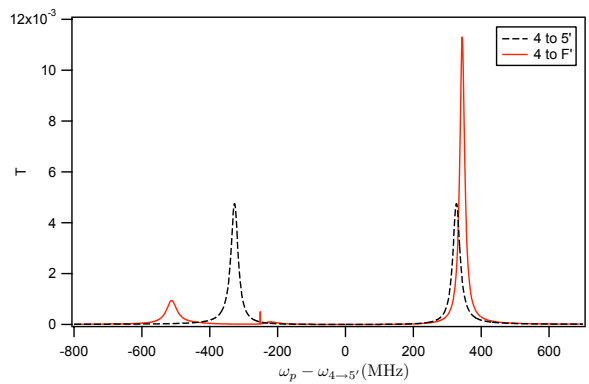}
	\caption[Effect of multiple hyperfine states on spectrum.]{\label{toroid_4_5} Normalized transmission $T$ of a cavity coupled to the $F=4\to F'= 5'$ transitions (dashed) and to the $F=4\to F'=\{3',4',5'\}$ transitions (red).  Parameters are the same as in Fig.~\ref{toroidplot}, with the cavity frequency fixed at $\omega_c = \omega_{4\to5'}$.}
\end{figure}

Fig.~\ref{toroid_4_5} demonstrates the importance of incorporating multiple hyperfine levels into the model of the atom when calculating the cavity transmission for the large values of $g$ expected in upcoming experiments \cite{toroid}.  The solid red curve denotes the transmission $T$ from Fig.~\ref{toroidplot} for a cavity fixed to be resonant with the $F=4\to F'=5'$ transition.  The dashed black curve indicates the transmission calculated using a model of the atom which includes all Zeeman states of the $F=4$ and $F'=5'$ manifolds, but no other hyperfine levels.  The substantial differences between the curves indicates that although the other hyperfine transitions are not resonant, the large coupling $g$ causes these transitions to have a significant effect on the atom-cavity system.

\section{Coupling to the Entire D2 Transition}
\label{sec:pbg}

\begin{figure}
	\includegraphics[width=3.3in]{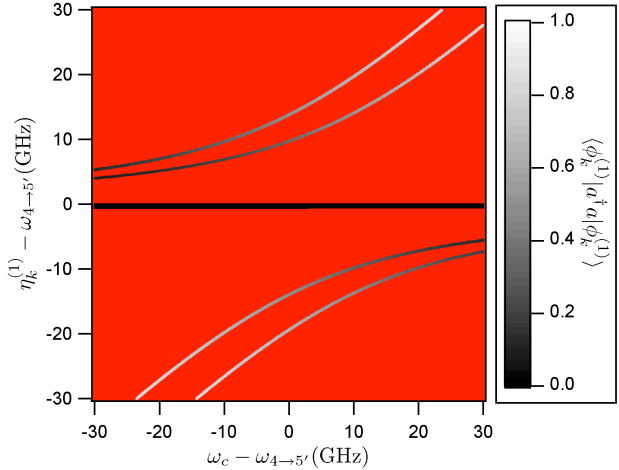}
	\caption[Eigenvalues of a single-mode cavity coupled to the Cs D2 transition.]{\label{pbg_eigen} Eigenfrequencies of excitations in the first manifold $\{\eta_k^{(1)}\}$ versus cavity detuning.  Color represents $\langle a^{\dag}a\rangle$ for each corresponding eigenvector.  The coupling $g=17$~GHz \cite{pbg}.  Although there are 48 eigenvectors in the manifold, there are only 27 unique eigenvalues due to symmetry.  These eigenvalues are clustered in 5 bands.}
\end{figure}

\begin{figure}
	\includegraphics[width=3.2in]{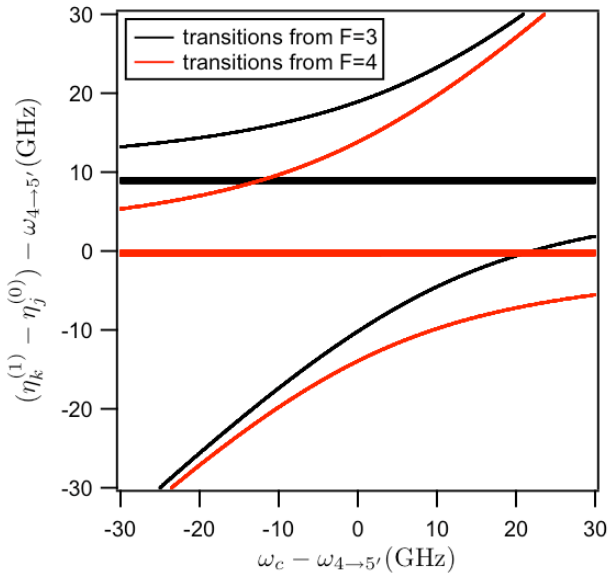}
	\caption[Transition frequencies of a single-mode cavity coupled to the Cs D2 transition.]{\label{pbg_levels} Difference frequencies for allowed single-quanta transitions between ground and first excited states, $\eta_k^{(1)}-\eta_j^{(0)}$, versus cavity detuning.  Eigenfrequencies are the same as in Fig.~\ref{pbg_eigen}.  Black (red) lines indicate transitions from ground states with the atom in the $F=3$ ($F=4$) manifold.
	}
\end{figure}

Now we will turn to the regime where $g$ is larger than both the ground- and excited-state hyperfine splittings.  This case is applicable for the expected parameters of cavity QED with photonic band gap cavities \cite{pbg}.  In this regime, the cavity mode couples to both ground-state hyperfine manifolds, and the Hamiltonian of the atom-cavity system in the absence of a driving field can be written
\begin{eqnarray}
\label{H_2}
H_2 &=&\sum_{F'}\omega_{F'}|F'\rangle \langle F'| - \omega_{GSS} |F=3\rangle\langle F=3| + \omega_c a^{\dag} a \nonumber \\
&+& g\sum_{F,F'}\Big( a^{\dag}D_{0}(F,F') + D_{0}^{\dag}(F,F')a \Big)
\end{eqnarray}
As we did earlier for $H_0$, we find the eigenvalues and eigenvectors of this Hamiltonian determined by the condition $H_2|\phi_k^{(N)}\rangle = \eta_k^{(N)} |\phi_k^{(N)}\rangle$.  In Fig.~\ref{pbg_eigen}, we plot the the frequencies $\eta_k^{(1)}$ of the lowest lying excitations, as well as how ``cavity-like'' the corresponding eigenmodes are, $\langle\phi_k^{(1)}| a^{\dag}a |\phi_k^{(1)}\rangle$.  The eigenvalues in the first excitation manifold separate into five bands.  The lowest and second-highest of these bands have eigenstates which are superpositions of $F=3$ atomic ground states with one photon in the cavity and $F'=\{2',3',4'\}$ atomic excited states with zero photons; the highest and second-lowest bands have eigenstates which are superpositions of $F=4$ states with one photon and $F'=\{3',4',5'\}$ states with zero photons.  

The central band is occupied by eigenstates the composition of which is dominated by atomic excited states. In particular, these eigenstates have a greatly suppressed coupling to the cavity mode as a result of quantum interference between transition amplitudes from atomic excited states with the same $m_F$ number but different values of $F'$. Similarly, with the assumption of a common reservoir for atomic spontaneous 
emission from the various hyperfine states (see below), 
these eigenstates also exhibit strongly suppressed spontaneous emission via $\pi$-polarized dipole transitions.  It should be noted that coupling to the D1 transition ($6S_{1/2} \to 6P_{1/2}$) does not result in a similar set of eigenstates with suppressed coupling; the absence of $F'=2',5'$ states precludes the possibility of the
required destructive quantum interference between $\pi$-polarized transitions.

\begin{figure}
	\includegraphics[width=3.3in]{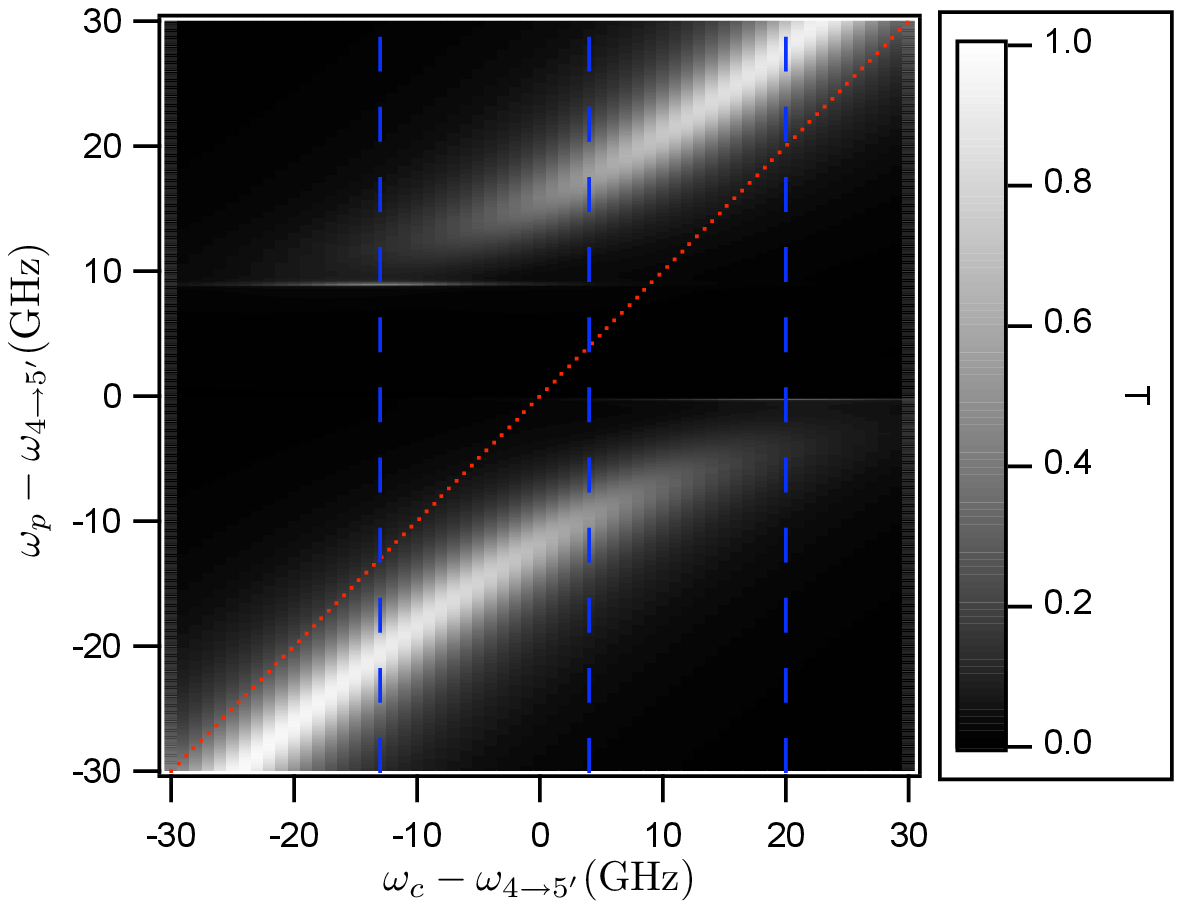}
	\caption[Predicted spectra of a PBG cavity coupled to the Cs D2 line versus cavity detuning.]{\label{pbgplot} Transmission (intracavity photon number, normalized to the empty cavity on resonance) versus probe and cavity detunings in the weak-driving limit (calculated in a Fock basis of \{0,1\} photons) from Eq.~\ref{master_2}.  The red dotted line indicates resonance for the uncoupled cavity.  The blue dashed lines indicate the cavity detunings in Figs.~\ref{pbg_slice_m13}, \ref{pbg_slice_p20} and \ref{pbg_slice}.  Rates are $(g,\kappa,\gamma)=(17,4.4,0.0026)$GHz~\cite{pbg},  $\varepsilon = 100\kappa\gamma/g$.  Values were computed on a grid with a $1$GHz ($0.05$GHz) spacing in $\omega_c$ ($\omega_p$).
	}
\end{figure}

We expect high transmission when a probe is tuned to be resonant with a transition from a ground state of the atom-cavity system to one of the states in the first excitation manifold.  The eigenvalues of the ground states are $\eta^{(0)}=0$ for states with the atom in the $F=4$ manifold and $\eta^{(0)}=-\omega_{GSS}$ for states with the atom in $F=3$.  In Fig.~\ref{pbg_levels}, we plot the difference frequencies for transitions between ground and first excited states, $\eta_k^{(1)}-\eta_j^{(0)}$, where $k,j$ are restricted to single-quantum transitions that can be excited by the cavity probe.  
Notice that although the eigenvalues of the Hamiltonian do not cross, the differences of eigenvalues between the ground and first excitation manifolds do have crossings.  These crossings correspond to a dual resonance condition, in which a transition from one hyperfine ground state to an excited state is resonant with a transition from the other hyperfine ground state to a different excited state. As we will show in a moment, this can lead to some distinctive features in the probe transmission spectrum.

\begin{figure}
	\includegraphics[width=3.3in]{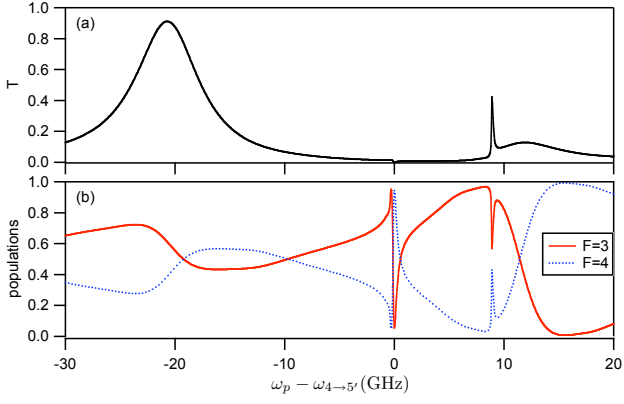}
	\caption[Transmission spectrum of a PBG cavity tuned to the lower dual resonance.]{\label{pbg_slice_m13} Normalized transmission $T$ (a) and atomic populations in the hyperfine ground states (b) as a function of probe detuning, with the cavity frequency fixed at $\omega_c = \omega_{4\to5'}-13$GHz.  Parameters are the same as in Fig.~\ref{pbgplot}. Note that because the cavity drive is weak, nearly all of the population is in the ground states.}
\end{figure}

We will now calculate the steady state of the driven, damped system. We will consider the cavity to be driven by a single coherent-state field at the frequency $\omega_p$.  Since the cavity mode can couple to all of the atomic ground states, a repump field is not needed. The Hamiltonian of the driven atom-cavity system under the rotating wave approximation, in the frame rotating with the probe, is
\begin{eqnarray}
\label{H_3}
H_3 &=&\sum_{F'}\Delta_{F'}|F'\rangle \langle F'| - \omega_{GSS} |F=3\rangle\langle F=3| + \Delta_c a^{\dag} a \nonumber \\
&+& g\sum_{F,F'}\Big( a^{\dag}D_{0}(F,F') + D_{0}^{\dag}(F,F')a \Big)\nonumber \\
&+&\varepsilon a^{\dag} +\varepsilon^{*} a ,
\end{eqnarray}
and the master equation for the evolution of the density matrix is
\begin{eqnarray}
\label{master_2}
\dot{\rho} = -i[H_3,\rho] + 
\kappa \mathcal{D}[a]\rho +
\gamma \sum_{q}\mathcal{D}\Big[\sum_{F,F'}D_{q}(F,F')\Big]\rho .
\end{eqnarray}
Note that in this limit, in which level shifts produced by the atom-field coupling may yield transitions
of similar frequencies to and from {\it different} hyperfine ground states (i.e., $F=3$ and $F=4$),
we assume that all atomic decays of a given polarization are into a common reservoir (without regard for the initial $F'$ and final $F$) \cite{cardimona83}.  
From the master equation (\ref{master_2}), 
we find the steady-state density matrix $\rho_{ss}$ and the steady-state normalized transmission $T$, plotted versus probe and cavity detunings in Fig.~\ref{pbgplot}.

\begin{figure}
	\includegraphics[width=3.3in]{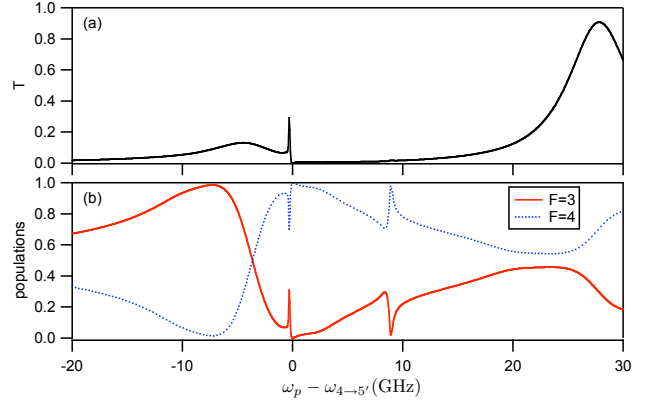}
	\caption[Transmission spectrum of a PBG cavity tuned to the upper dual resonance.]{\label{pbg_slice_p20} Normalized transmission $T$ (a) and atomic populations in the hyperfine ground states (b) as a function of probe detuning, with the cavity frequency fixed at $\omega_c = \omega_{4\to5'}+20$GHz.  Parameters are the same as in Fig.~\ref{pbgplot}. Note that because the cavity drive is weak, nearly all of the population is in the ground states.}
\end{figure}  

These transmission spectra reflect the structure of the eigenvalues plotted in Fig.~\ref{pbg_levels}, although since $\kappa$ is not substantially smaller than $g$ for the parameter set considered, the correspondence is perhaps not as pronounced as for the previous section.
While the transmission spectra are dominated by a pair of broad peaks with widths of the order of $\kappa$, of particular interest are sharp features at $\omega_p \approx \omega_{4\to5'}-0.3$~GHz and $\omega_p \approx \omega_{4\to5'}+8.9$~GHz.  These transmission features are particularly strong at the cavity detunings where transition frequencies of $H_2$ cross, i.e., where the dual resonance condition is satisfied, which for the parameters $(g,\kappa,\gamma)=(17,4.4,0.0026)$GHz occurs at $\omega_c \approx \omega_{4\to5'}+20$~GHz and $\omega_c \approx \omega_{4\to5'}-13$~GHz.  The steady-state transmission at these cavity detunings is plotted in Figs.~\ref{pbg_slice_m13}(a) and \ref{pbg_slice_p20}(a) versus probe detuning.

Also plotted, in Figs.~\ref{pbg_slice_m13}(b) and \ref{pbg_slice_p20}(b), are the total populations in the $F=3$ and $F=4$ ground-state manifolds, which illustrate that
the sharp peaks in cavity transmission are associated with significant optical pumping effects. 
For the case illustrated in Fig.~\ref{pbg_slice_p20},
the transitions which satisfy the dual resonance condition are between the $F=3$ ground-state manifold and a manifold of excited eigenstates which have a significant photon component, and between the $F=4$ ground-state manifold and the central band of atom-like eigenstates. Weak dissipative channels (primarily atomic spontaneous emission) can transfer population between the two transitions in a manner that depends sensitively on the probe field detuning and the atomic state compositions of the excited eigenstates. 
Pronounced optical pumping effects between the different $m_F$ levels also occur as the probe field is tuned to the various atom-like eigenstates as
a result of the suppression of $\pi$-polarized spontaneous emission from each of 
these states.

\begin{figure}
	\includegraphics[width=3.3in]{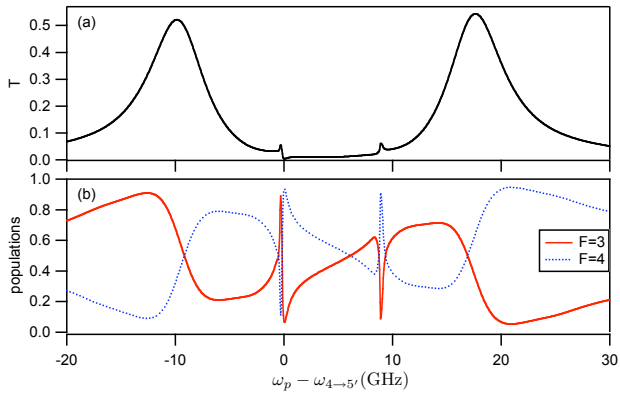}
	\caption[Spectrum and ground-state populations of Cs coupled to a PBG cavity.]{\label{pbg_slice} Normalized transmission $T$ (a) and atomic populations in the hyperfine ground states (b) as a function of probe detuning, with the cavity frequency fixed at $\omega_c = \omega_{4\to5'}+4$GHz.  Parameters are the same as in Fig.~\ref{pbgplot}. Note that because the cavity drive is weak, nearly all of the population is in the ground states.}
\end{figure}

In Fig.~\ref{pbg_slice}(a), we plot the normalized steady-state transmission versus probe detuning with the cavity frequency fixed between the frequencies of the $F=3\to F'$ and $F=4\to F'$ transitions.  Two small narrow peaks associated with the dual resonance effect are still apparent, and the atomic populations, plotted in Fig.~\ref{pbg_slice}(b), now show very strong and abrupt pumping into the $F=3$ or $F=4$ manifolds around these peaks.

Future experiments with single atoms coupled to photonic band gap cavities should be able to study these sharp features.  They should be relatively easy to measure because although they are narrow in probe frequency (which is easily controlled), they are robust against changes in cavity frequency (which is harder to control experimentally) of the order of $\kappa$. 

\section{Conclusion}
We have presented results of the calculation of the weak-field steady-state transmission of a single-mode linearly polarized optical resonator coupled to the D2 transition of a single Cesium atom.
Our results are for a regime of single-photon dipole coupling strength not previously 
considered, but of relevance to planned experiments with microtoroid and photonic bandgap cavities, as well as with other recently-implemented atom-chip microcavity systems \cite{treutlein06,barclay06}. 
They necessarily take into account the entire atomic hyperfine
structure and comparison with simpler models highlights the importance of doing so.
In addition to features expected from a strongly coupled atom-cavity system, they also
reveal interesting and significant quantum interference phenomena associated with 
the coupling of different atomic transitions to the same mode or modes of 
the electromagnetic field.

\begin{acknowledgments}
This research is supported by the National Science Foundation, by the Caltech MURI Center for Quantum Networks, and by the Advanced Research and Development Activity (ARDA). ASP acknowledges support from the Marsden Fund of the Royal Society of New Zealand.
\end{acknowledgments}

\bibliography{hyperfine}

\end{document}